\documentclass[10pt,journal,compsoc]{IEEEtran} \IEEEoverridecommandlockouts

\ifCLASSOPTIONcompsoc
  \usepackage[nocompress]{cite}
\else
  \usepackage{cite}
\fi

\ifCLASSINFOpdf
  \usepackage[pdftex]{graphicx}
\else
\fi
\usepackage[utf8]{inputenc}
\usepackage[english]{babel}
\usepackage{graphicx}
\usepackage[group-separator={,}]{siunitx}
\usepackage{pifont}
\newcommand{\cmark}{\ding{51}}%
\newcommand{\xmark}{\ding{55}}%
\usepackage{numprint}
\usepackage[T1]{fontenc}
\usepackage{lipsum}
\usepackage{stfloats} 
\usepackage{array}
\usepackage[inline]{enumitem}
\usepackage{color}
\usepackage{hyperref}
\usepackage{listings}
\usepackage{xcolor}
\usepackage{csquotes}
\usepackage{amsmath}
\usepackage{xspace}
\usepackage{acronym}
\usepackage{xstring}
\usepackage{amsmath,amsthm}
\usepackage[nobreak]{mdframed}
\usepackage{framed}
\usepackage{multirow}
\usepackage{float}

\ifCLASSOPTIONcompsoc
  \usepackage[nocompress]{cite}
\else
  \usepackage{cite}
\fi

\usepackage{macros}
\usepackage{keywords}
\usepackage{etoolbox,xpatch}

\patchcmd{\thmhead}{(#3)}{#3}{}{}
\xpatchcmd{\@thm}{\fontseries\mddefault\upshape}{}{}{}

\newcolumntype{L}{>{\centering\arraybackslash}m{3cm}}

\makeatother
\theoremstyle{definition}
\theoremstyle{definition}

\acrodef{URL}{Uniform Resource Locator}
\acrodef{JNDI}{Java Naming and Directory Interface}
\acrodef{JVM}{Java Virtual Machine}
\acrodef{SBOM}{Software Bill of Materials}
\acrodef{JMH}{Java Microbenchmark Harness}
\acrodef{SGX}{Intel Software Guard Extensions}
\acrodef{eBPF}{Extended Berkeley Packet Filter}
\acrodef{JSM}{Java Security Manager}
\acrodef{IMA}{Integrity Measurement Architecture}
\acrodef{BOMI}{Bill of Material Index}
\acrodef{CVE}{Common Vulnerabilities and Exposures}
\newcommand{\rqscale}{RQ1: What is the scale of \ac{BOMI} for all the study subjects?\xspace}
\newcommand{\rqeffectiveness}{RQ2: To what extent can \toolname mitigate high-profile attacks in Java?\xspace}
\newcommand{\rqapplicability}{RQ3: Is \toolname compatible with real-world applications?\xspace}
\newcommand{\rqperformance}{RQ4: What is the overhead of \toolname?\xspace}
\newcommand{\toolname}{\textsc{SBOM.exe}\xspace}
\newcommand{\bomi}{\ac{BOMI}\xspace}
\newcommand{\bomienvironment}{\ac{BOMI}-Environment\xspace}
\newcommand{\bomiruntime}{\ac{BOMI}-Runtime\xspace}
\newcommand{\bomisupplychain}{\ac{BOMI}-SupplyChain\xspace}
\let\oldtexttt\texttt
\renewcommand{\texttt}[1]{\oldtexttt{\StrSubstitute[0]{#1}{.}{.\allowbreak}}}

\newcommand{\GHilight}{\makebox[0pt][l]{\color{green}\rule[-4pt]{1\linewidth}{10pt}}}
\newcommand{\RHilight}{\makebox[0pt][l]{\color{pink}\rule[-4pt]{1\linewidth}{10pt}}}

\definecolor{javared}{rgb}{0.6,0,0} 
\definecolor{javagreen}{rgb}{0.25,0.5,0.35} 
\definecolor{javapurple}{rgb}{0.5,0,0.35} 
\definecolor{javadocblue}{rgb}{0.25,0.35,0.75} 

\lstdefinestyle{diff}{
    escapechar=\%
}

\lstdefinelanguage{CallStack}{
  morekeywords={},
  sensitive=false,
  morecomment=[l]{//},
  morestring=[b]"
}

\lstdefinestyle{customCallStack}{
  numbers=none,
  language=CallStack,
  commentstyle=\color{gray},
  basicstyle=\ttfamily\footnotesize,
  breaklines=true,
  rulecolor=\color{black},
  captionpos=b,
  belowcaptionskip=\baselineskip,
  moredelim=[is][\color{blue}]{~}{^},
  moredelim=**[is][\color{black}]{^}{^},
  moredelim=**[is][\color{red}]{^}{\$},
  moredelim=**[is][\color{black}]{\$}{)},
}

\begin{document}

\title{\toolname: Countering Dynamic Code Injection based on Software Bill of Materials in Java}

\author{
  Aman Sharma,
  Martin Wittlinger,
  Benoit Baudry,
  Martin Monperrus

  \IEEEcompsocitemizethanks{
  \IEEEcompsocthanksitem A. Sharma and M. Monperrus are with the KTH Royal Institute of Technology, Stockholm, Sweden\protect\\
  Email: \{amansha, monperrus\}@kth.se\protect\\
  \IEEEcompsocthanksitem M. Wittlinger is with the HDI Group, Cologne, Germany\protect\\
  Email: martin.wittlinger@hdi.de\protect\\
  \IEEEcompsocthanksitem B. Baudry is with the Universtit\'e de Montr\'eal, Montr\'eal, Canada\protect\\
  Email: benoit.baudry@umontreal.ca
}
}

\IEEEtitleabstractindextext{%
  \begin{abstract}
    Software supply chain attacks have become a significant threat as software development increasingly relies on contributions from multiple, often unverified sources.
    The code from unverified sources does not pose a threat until it is executed.
    \texttt{Log4Shell} is a recent example of a supply chain attack that processed a malicious input at runtime, leading to remote code execution.
    It exploited the dynamic class loading facilities of Java to compromise the runtime integrity of the application.
    Traditional safeguards can mitigate supply chain attacks at build time, but they have limitations in mitigating runtime threats posed by dynamically loaded malicious classes.
    This calls for a system that can detect these malicious classes and prevent their execution at runtime.
    
    This paper introduces \toolname, a proactive system designed to safeguard Java applications against such threats.
    \toolname constructs a comprehensive allowlist of permissible classes based on the complete software supply chain of the application.
    This allowlist is enforced at runtime, blocking any unrecognized or tampered classes from executing.
    We assess \toolname's effectiveness by mitigating 3 critical CVEs based on the above threat.
    We run our tool with 3 open-source Java applications and report that our tool is compatible with real-world applications with minimal performance overhead.
    Our findings demonstrate that \toolname can effectively maintain runtime integrity with minimal performance impact, offering a novel approach to fortifying Java applications against dynamic classloading attacks.
    \\
    \textbf{Publicly-available repository} - \url{https://github.com/chains-project/sbom.exe}
  \end{abstract}

  }

\maketitle

\IEEEdisplaynontitleabstractindextext

%
\IEEEpeerreviewmaketitle

\section{Introduction}
\label{sec:introduction}

Developers reuse a lot of third-party dependencies~\cite{cox_surviving_2019,soto-valero_multibillion_2022,raemaekers_analysis_2015} to build software applications.
The process of building the application using the set of all dependencies is known as the software supply chain of an application.
While this practice is good as it avoids reinventing the wheel~\cite{krueger_software_1992}, it is challenging for developers to keep track of the reliability, maintainability and security of their software supply chain~\cite{balliu_challenges_2023}.
In the latter case, it has recently been observed, with high-profile attacks, that third-party libraries can be exploited by malicious actors, leading to so-called ``software supply chain attacks''~\cite{ladisa_taxonomy_2022,ohm_backstabber_2020,enck_top_2022}.

Software supply chain attacks are a significant threat to the software security landscape as acknowledged by ENISA~\cite{noauthor_enisa_nodate} and the White House~\cite{house_executive_2021}.
In 2023, there were twice as many software supply chain attacks as in 2019-2022 combined~\cite{sonatype_2023_nodate}.
A few reactive and proactive techniques have been proposed in the literature to mitigate software supply chain attacks~\cite{ladisa_taxonomy_2022}.
Reactive techniques are based for example on bots~\cite{mohayeji_investigating_2023} to update dependency regularly and tools to scan for \ac{CVE}s~\cite{imtiaz_comparative_2021} in dependencies.
However, these only act after the exploit has been discovered and reported.
The main proactive technique to help prevent software supply chain attacks is to systematically create \ac{SBOM}. In a nutshell, an \ac{SBOM} is a complete list of dependencies and tools used to build a software application.
\ac{SBOM}s increase transparency in the software supply chain, and are good for accountability~\cite{house_executive_2021}, but they cannot prevent attacks at runtime.

In this paper, we propose a novel technique to mitigate a class of software supply chain attacks based on code injection in third-party dependencies at runtime~\cite{ohm_backstabber_2020}.
This type of attack received major attention when the high-profile attack \texttt{Log4Shell} (CVE-2021-44228~\cite{noauthor_nvd_nodate}) was released, demonstrating that the dependency \texttt{Log4j}, which is part of millions of Java applications' software supply chain~\cite{turunen_log4shell_2021}, was vulnerable.
The attack was possible because the dependency had a vulnerability enabling remote code execution when processing a malicious input.
More importantly, the attack was entirely based on dynamic class loading facility that exists in Java.

Conceptually, if an attacker knows that a dependency uses Java dynamic features, then they can exploit these features to compromise any dependent application.
The solution is obviously not to forbid these dynamic features, because they are used in virtually all mission and business critical applications, incl. all Spring Java web apps.
Also, neither \ac{SBOM}, reproducible builds, code signing, nor version pinning would prevent these attacks as the malicious code in these attacks appears only while the application is running.

We propose \toolname, a system that ensures the integrity of the Java runtime, by monitoring the software supply chain of an application at runtime, in order to detect and prevent the execution of injected malicious code. 
\toolname works by building a comprehensive allowlist of classes that are allowed to be executed.
We call this list the \ac{BOMI}.
This first task is hard due to the widespread usage of dynamically generated code in Java applications. 
Second, the \ac{BOMI} contains the checksums of all allowed classes, both from the application and its supply chain.
This is also hard because runtime code generation may be non-deterministic, and differs because of various reasons, incl. Java version~\cite{xiong_build_2022}.
To overcome this problem and make the \bomi robust, we propose a novel technique of bytecode canonicalization that mitigates all sources of non-deterministic features in Java bytecode in order to compute reliable and secure checksum.
Then, \toolname ensures that no unknown class is executed at runtime by comparing the checksum of the class about to be loaded with the reference checksum that is in the \bomi.
Finally, and most importantly, the tool stops the execution if it intercepts any unknown or tampered class.
To our knowledge, \toolname is the first system that reasons and embeds the software supply chain at runtime in order to block malicious code injection in Java.

We evaluate the effectiveness of \toolname by testing it against 3 real-world, critical vulnerabilities in Java libraries: \texttt{Log4j}~\cite{noauthor_nvd_nodate}, \texttt{H2}~\cite{noauthor_nvd_nodatea}, and \texttt{Apache Commons Configuration}~\cite{noauthor_nvd_nodateb}.
First, we replicate these 3 critical \ac{CVE}s in a lab setting, next, we show that \toolname fully mitigates them by stopping the execution of the application before the malicious class is executed.
We also demonstrate that our \toolname does not break real-world software by running 3 open-source Java applications with nominal workloads - \texttt{PDFBox}~\cite{noauthor_apache_nodate}, \texttt{Ttorrent}~\cite{petazzoni_mpetazzoni_2024}, and \texttt{GraphHopper}~\cite{noauthor_graphhopper_2017}.
Finally, our performance measurements with state-of-the-art microbenchmarking show that the overhead incurred by \toolname is atmost 1\% which is negligible.

The main contributions of this paper are:
\begin{itemize}
  \item \toolname, a novel system that ensures the integrity of the Java runtime against code injection. It works by computing a comprehensive allowlist of classes using the software supply chain of the application.
  \item A robust algorithm and tool to compute checksums of Java bytecode, removing non-deterministic features, and enabling sound malicious code detection.
  \item A series of experiments showing 1) the effectiveness of \toolname in mitigating real-world Java vulnerabilities including \texttt{Log4Shell}, 2) the compatibility of \toolname with existing applications, and 3) the absence of significant overhead.
  \item A publicly available tool and consolidated reproducible attacks on GitHub~\cite{sharma_sbom_2024} for future research on this topic.
\end{itemize}

\section{Background}

\subsection{Dynamicity of the Java Runtime}
\label{sec:dynamicity-java}
Java is a dynamic programming language, in the sense that it features different capabilities to load and modify code at runtime~\cite{noauthor_classloader_nodate}.
A classloader in Java is responsible for loading classes into the \ac{JVM}.
A typical classloader takes in the name of the class and finds binary code corresponding to it on disk.
Dynamic class loading does not require the binary code to exist on disk since the start of \ac{JVM}.
For example, it enables downloading binary code on the fly or generation of binary code at runtime.
In the following, we enumerate the main mechanisms through which a class can be loaded dynamically in Java.

First, classloaders can be extended to execute classes from a remote source~\cite{noauthor_urlclassloader_nodate} or compiled code at runtime~\cite{noauthor_javacompiler_nodate}.
Both of these APIs are open for public usage and extensively used by the native SDK.

Proxies~\cite{noauthor_proxy_nodate} are runtime generated classes that add common functionalities to some classes in the application.
For example, a proxy class can have functions to record the time taken for execution.
The code for proxies is generated at runtime.
Within Java, a major usage of proxies is to generate the code corresponding to Java annotations~\cite{noauthor_annotations_nodate}.

Java allows classes to introspect themselves.
This is called reflection in Java and this feature also leverages dynamic code generation.
Internally, Java creates subclasses of \texttt{MagicAccessorImpl}~\cite{noauthor_magicaccessorimpl_nodate} which grants access to the Java runtime to members of the class which otherwise would be inaccessible.
Introspection is an essential feature of frameworks.
For example, JUnit uses reflection to find the test methods in a class.

Finally, Java bytecode has an instruction called \texttt{invokedynamic}~\cite{evans_understanding_2021} that allows bootstrapping methods at runtime.
It bootstraps methods by dynamically generating a `hidden class'~\cite{chung_jep_2019} that contains the implementation of the method.
This instruction is used to implement lambda expressions in Java, which is essential in modern versions of Java.
Java records also rely on this feature to implement their members like \texttt{equals}.

To sum up, Java is a highly dynamic platform, and its dynamic features are foundational for most notable Java capabilities and usages. 
However, this dynamicity can be used for malicious purposes, which is the problem we address in this paper.
We will study real-world instances of such attacks in~\autoref{sec:threat-model} and~\autoref{sec:rq-applicability}.

\subsection{The Software Supply Chain}

Software Supply Chain refers to the sequence of steps and inputs resulting in the creation of a software artefact~\cite{noauthor_terminology_nodate}.
We define the important terms used in the paper regarding the software supply chain.

\emph{Build System.}
A build system is a piece of software that takes in the source code and all its dependencies to create a package or a software artefact.
Some examples of build systems are \texttt{mvn}, \texttt{npm}, and \texttt{pip}.

\emph{Package Registry.}
A package registry is a trusted central service that hosts packages, dependencies, or software components.
Packages are deployed to registries so that other developers can use them.
Some examples of package registries are Maven Central, the npm registry, and PyPI.

\emph{\ac{SBOM}.}
\ac{SBOM} -- software bill of material -- is a formal, machine-readable inventory of software components, information about those components, and their hierarchical relationships~\cite{ntia_sbom_2021}.
Its goal is to enable transparency of the software supply chain.
This enables multiple uses to multiple stakeholders~\cite{ntia_roles_2019}, such as license compliance analysis or vulnerability management.

One of the key features of \ac{SBOM} is to report the complete list of software components or dependencies of a software package~\cite{ntia_minimum_2021}.
The dependencies here refer to both direct and indirect dependencies.
The direct dependencies refer to the ones declared by the developers to leverage the additional functionalities.
The indirect dependencies are the ones that are required by the direct dependencies and are implicitly trusted by the developers.

\section{Threat Model}
\label{sec:threat-model}

Our goal is to protect against malicious usage of dynamic features in Java, which were presented in \autoref{sec:dynamicity-java}.
Recall that, in Java, a class can be downloaded from a remote source or generated at runtime~\cite[§3]{noauthor_chapter_nodate} and executed.
In most cases, the developer is unaware of those dynamic classes as they are unknown when they are writing and building the application.

The threat we are defending against is the loading of malicious classes by the Java runtime, which is considered a kind of code injection attack \cite[§"Network Class Loaders and Security Issues"]{mahmoud_understanding_2004}.
This class of attack has been studied in the literature \cite{holzinger_indepth_2016}.

In our context, we equate `injected code' with unknown classes.
These classes are neither part of the application, nor of its dependencies, nor the Java standard library.
Note that an unknown class can also be a modified version of a known class, triggered by malicious actor tampering with application classes.
Per the terminology of Holzinger et al.~\cite{holzinger_indepth_2016}, we mitigate against the vulnerability ``Loading of arbitrary classes''.

\begin{figure}
  \centering
  \includegraphics[width=\columnwidth]{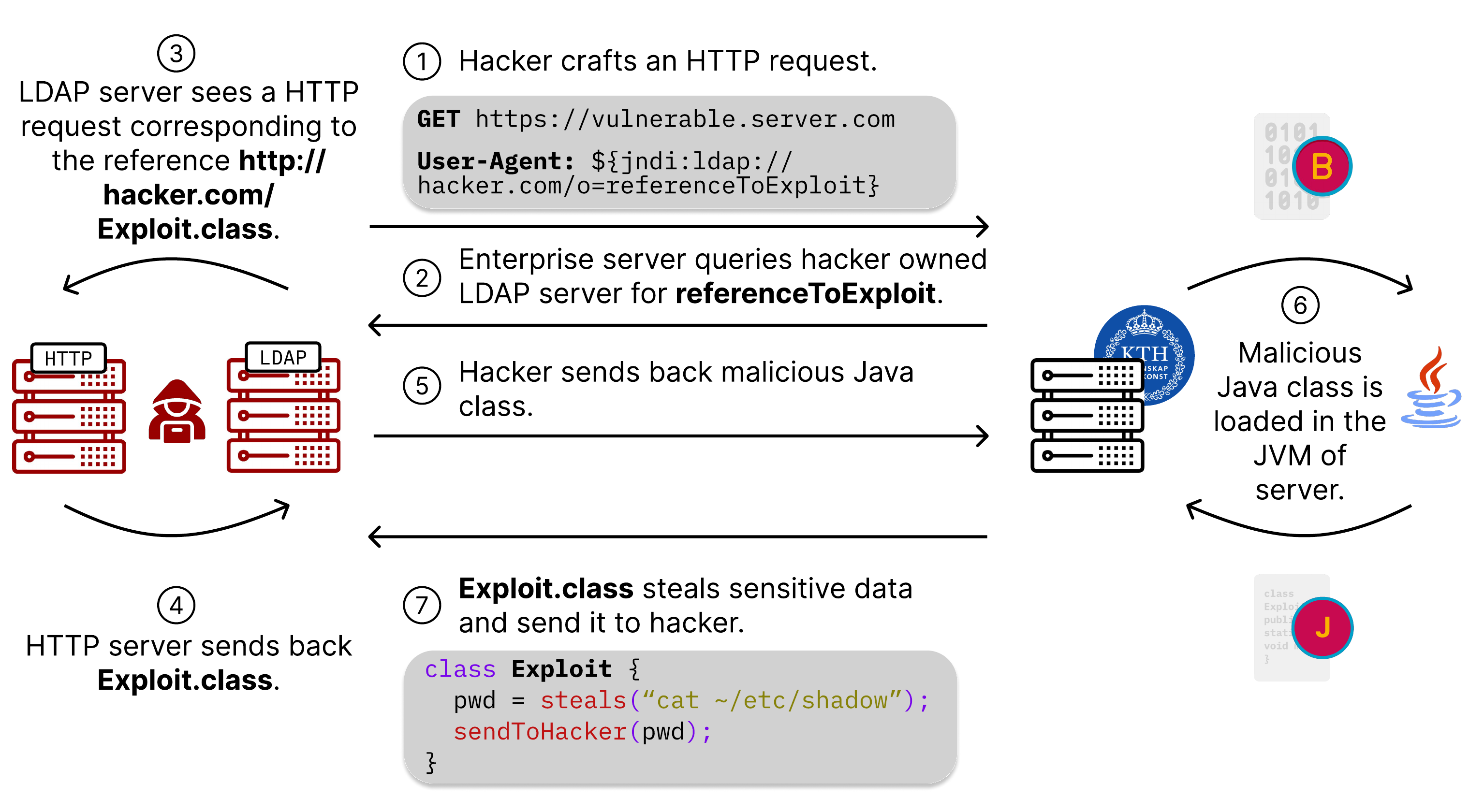}
  \caption{Overview of the Log4Shell vulnerability.}
  \label{fig:log4j}
\end{figure}

We will study real-world instances of these attacks in~\autoref{sec:rq-effectiveness}. For example, \texttt{Log4Shell}~\cite{noauthor_nvd_nodate} is one infamous example where unknown code is downloaded from a remote server and executed.
\autoref{fig:log4j} gives an overview of a \texttt{Log4Shell} attack.
The attacker, first, creates a malicious class file \texttt{Exploit.class} and hosts it on a server controlled by them.
The malicious class file is hosted on the HTTP server and the LDAP server stores a reference to the HTTP server via an HTTP URL.
Then, the attacker sends a crafted request with expression \texttt{\$\{jndi:ldap://hacker.com/o=referenceToExploit\}} in the header.
JNDI stands for Java Naming and Directory Interface and it allows one to look up resources in a directory.
The LDAP server with domain name \texttt{hacker.com} is used to store the reference to the HTTP server.
\texttt{http://hacker.com/Exploit.class} is queried and it returns the malicious class file.
The enterprise application is vulnerable to this as it is using \texttt{Log4j}.
\texttt{Log4j} logs the value in the User-Agent header.
This is normally done to record the telemetry data of the application.
However, in this case, the expression is interpreted and the \ac{JVM} running loads \texttt{Exploit.class}.
Finally, the malicious class can execute arbitrary offensive commands, e.g. to steal private data or perform ransomware attacks.

To sum up, in this paper, we propose a technique to mitigate malicious code execution in Java that exploits the dynamic features of the platform.

\section{Design \& Implementation}
\begin{figure*}
  \centering
  \includegraphics[width=\textwidth]{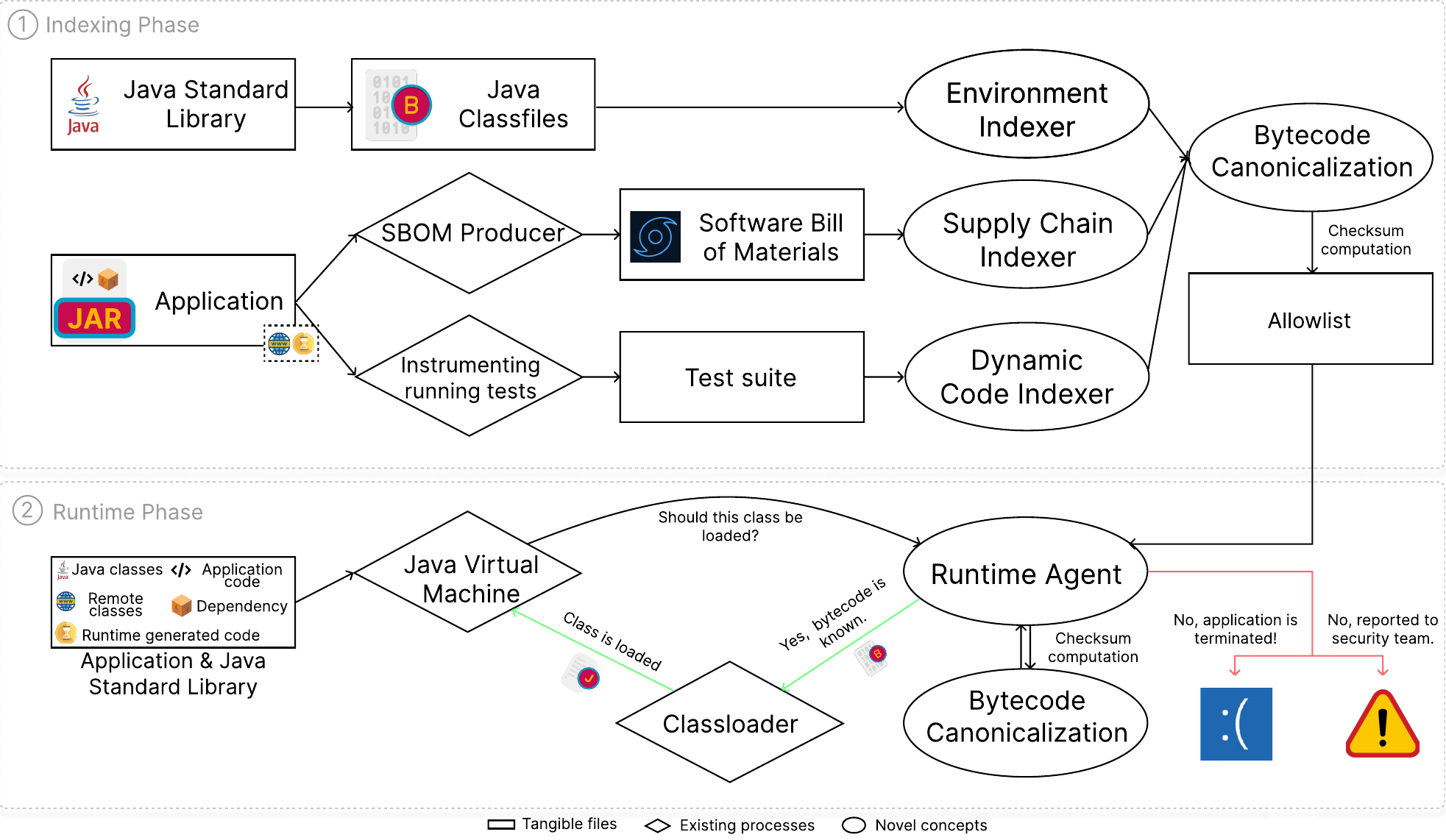}
  \caption{Overview of SBOM.exe, a novel system to detect and mitigate code injection attacks in Java systems.}
  \label{fig:overview}
\end{figure*}
We now present the design of \toolname, our novel system for ensuring supply chain integrity at runtime in Java.

\subsection{System Overview}

\autoref{fig:overview} presents an overview of  \toolname, a system that verifies the binary integrity of an application's dependencies at runtime.
The system operates in two different phases. 

At build time, the \emph{indexing phase} consists of building an allowlist of all binary classes that are allowed to be executed by the \ac{JVM} in production. We call this list the \texttt{BOMI}, which stands for Bill Of Material Index. Then at runtime, \toolname verifies that no unknown class is executed using the BOMI as the ground truth of accepted classes. The \texttt{SBOM Runtime Watchdog} is the key component at this stage.
It verifies the acceptability of a class by comparing the checksum of a canonical version of a Java class.
This checksum is computed exactly the same way in the indexing phase.

There are two major novel concepts in our system.

\textbf{\ac{BOMI}.} 
A Bill of Material Index, shortened as `BOMI', is an allowlist of binary classes that are allowed to be executed by the \ac{JVM}.
The core idea behind maintaining an allowlist is to restrict the execution of classes that are not known by the developers.
\toolname fully automates this process of indexing all the classes, not requiring the developer to do any manual work.
The BOMI captures the identity of allowed classes by computing a checksum over a canonical version of Java bytecode (\autoref{sec:bytecode-canonicalization}).

\textbf{\ac{SBOM} Runtime Watchdog.}
The goal of SBOM Runtime Watchdog is to ensure that no unknown class is executed.
It enforces the \ac{BOMI} at runtime by controlling the loading of classes.
In other words, it checks for equivalence of the checksum of the class to load with the checksum of the class in the \ac{BOMI}.
Since some classes are non-deterministically generated at runtime, we canonicalize (\autoref{sec:bytecode-canonicalization}) them to ensure that the checksum equivalence is correct.

In the \texttt{Indexing} phase (top part of \autoref{fig:overview}), \toolname lists all Java classes necessary to run the application before it is deployed to production.
In Java, default class loaders allow loading classes from filesystem~\cite[§2]{noauthor_classloader_nodate}.
These can be used to load environment classes and supply chain classes.
The default class loaders can be extended to download classes from a remote source or generate classes at runtime.
We refer to those classes as dynamically loaded classes.

Now, we define each of the three sets of classes.
First, environment classes are the built-in classes provided by the Java standard library.
Second, supply chain classes are classes written by the developers of the application, as well as the classes declared as dependencies to the application.
Finally, dynamically-loaded classes are classes that only appear at runtime. They are either downloaded from a remote source or generated.

The bottom part of \autoref{fig:overview} is about the \texttt{Runtime} phase of \toolname.
In production, a JVM is spawned to run an application.
We propose to attach an \emph{SBOM Runtime Watchdog} to the JVM.
The SBOM Runtime Watchdog is attached to the application during startup with the Java agent mechanism.
It takes as input the \ac{BOMI}, which will be used to detect unknown or tampered binary classes.
To sum up, for a class to be accepted and used in production, it must be present in the \ac{BOMI}, and must have the same canonical checksum as recorded in the \texttt{Indexing} phase.

\subsection{Indexing}

The core idea in the \texttt{Indexing} phase is to list all Java classes necessary to run the application.

\toolname has three indexing components: environment, supply chain, and dynamic code indexer.
All these processes execute sequentially and write the classes to the \ac{BOMI}.
Thus, the BOMI is the union of all classes obtained from the three indexers.
We now describe each of the indexers in detail.

\subsubsection{Environment Indexer}
The environment index contains information about the internal classes of the runtime environment, Java in our paper.
\toolname builds this index by scanning the Java standard library and recording the checksum of each standard class in the \ac{BOMI}.
This forms the first part of the \ac{BOMI} which is referred to as \bomienvironment.

We rely on ClassGraph~\cite{hutchison_classgraph_2024} which has APIs to scan the Java standard library.
It locates all Java classes packaged inside the Java distribution and forwards them to the environment indexer for checksum computation.
This index is generated statically and needs to be updated only when the Java version of the application changes.
The only requirement of this process is to have Java installed on the system.
The index always contains the same classes for a given Java version and operating system and its generation is reproducible.

\autoref{lst:environment-index} shows an excerpt from the environment index.
Each line of the excerpt contains a map of a class from the Java standard library to the checksum of its bytecode.
The map helps to quickly look up the checksum of a class based on its name.
Note that this is an excerpt and the actual index can contain classes in the order of tens of thousands.
Although not all of them are used by the application, we record all of them so that we can guard against classes masking as internal Java standard library classes.
For example, a malicious actor can create a class \texttt{jdk.internal.MaliciousClass} that could be downloaded at runtime. It could be deemed safe if we don't record all the internal classes of the Java standard library.

\begin{lstlisting}[backgroundcolor=\color{gray!10}, caption={Excerpt from the environment index of the Java standard library.}, label=lst:environment-index,language=JSON,belowskip=-0.4\baselineskip]
  {"java/lang/ClassLoader":{"checksum":"041c0b7"}}
  {"java/lang/Exception":{"checksum":"50c1c7d"}}
  {"java/lang/String":{"checksum":"5f158f2"}}
  {"java/lang/Math":{"checksum":"d3210a9"}}
  {"java/lang/System":{"checksum":"89b5804"}}
  ...
\end{lstlisting}

\subsubsection{Supply Chain Indexer}

The supply chain index is created from the \ac{SBOM} of the application.
The \ac{SBOM} is taken as input from the \ac{SBOM} producer.
An \ac{SBOM} maps all the classes in the supply chain of the application to a dependency in a trusted package registry.
This step is outside the threat model since the production of \ac{SBOM} is carried out by the developer of the dependency who uploads it to the trusted package registry.
From the trusted package registry, \toolname downloads a JAR file for each dependency.
The indexer extracts the JAR file and analyzes all the classes.
Finally, the indexer computes the checksum and writes the class name and checksum to the \ac{BOMI}.
This forms the second part of the \ac{BOMI} which is referred to as \bomisupplychain.

An excerpt from the supply chain index of a Java application is shown in \autoref{lst:supply-chain-index}.
This Java application has a single class that performs logging operations with \texttt{Log4j} dependency.
Hence, the first line in the excerpt is the class written by the author of the application.
The next three classes are from the \texttt{Log4j} dependency.
The last class is from the \texttt{Log4j} dependency but has two checksums.
This is because \texttt{Log4j} is a multi-release JAR and some classes can have different  implementations based on the Java version.

\begin{lstlisting}[backgroundcolor=\color{gray!10},caption={Excerpt from the supply chain index in the \ac{BOMI}.}, label=lst:supply-chain-index,language=JSON,belowskip=-0.4\baselineskip]
  {"org/example/Main":[{"checksum":"f8423f5"}]}
  {"org/apache/logging/log4j/core/lookup/JndiLookup":[{"checksum":"dacd441"}]}
  {"org/apache/logging/log4j/LogManager":[{"checksum":"4dafd50"}]}
  {"org/apache/logging/log4j/core/Logger":[{"checksum":"85a7729"}]}
  {"org/apache/logging/log4j/util/StackLocator":[{"checksum":"0cd3eb5"}, {"checksum":"2d62281"}]}
  ...
\end{lstlisting}

\subsubsection{Dynamic Code Generation Indexer}

Most Java applications~\cite{fourtounis_static_2018},\cite{landman_challenges_2017} depend on classes that are generated during runtime or are downloaded from a remote source.
By construction, those classes are unknown statically and are not indexed by the environment and supply chain indexers.
In \toolname, we take special care of supporting binary classes that are downloaded or generated at runtime.
The dynamic code generation indexer is responsible for tracing them and for generating the dynamic code index.
This index contains unique information about all classes that are neither part of the supply chain nor the Java standard library.
This forms the third and final part of the \ac{BOMI} which is referred to as \bomiruntime.

Let us delve into dynamic code generation in Java.
Internally, Java uses five mechanisms to generate runtime classes: proxy classes, classes generated using CGLIB, annotation, reflective invocation, and lambda expressions \cite{ba_rim4j_2018}.
Also, developers use libraries like ByteBuddy~\cite{noauthor_byte_nodate}, ASM~\cite{noauthor_asm_nodate}, and Javassist~\cite{noauthor_javassist_nodate} to generate classes at runtime.

To capture these classes, \toolname runs the tests of the application and records all classes that are downloaded or generated during test execution.
This assumes that the application has a good test suite that exercises the core functionalities of an application.
Our experiment (see \autoref{sec:rq-applicability}) demonstrates that this assumption does not miss any dynamic code that is required to run real-world applications in production.

The dynamic code indexer records the checksum of each class dynamically loaded or generated during test suite execution.
We show an excerpt from the dynamic code index in \autoref{lst:dynamic-code-index}.
All the classes are generated at runtime.
The first class is generated to implement annotations in Java.
Since annotations are defined by the developer as interfaces, Java generates implementations of them at runtime.
The second class is generated by the Java standard library to get reflective access to the constructor.
This is needed by Java for garbage collection.
The third class is generated to store argument values for methods that are invoked later in the runtime.
The fourth class is generated by the Nashorn JavaScript engine to evaluate the JavaScript code.
The last class is an example of a class generated by a Java framework to manage server configuration.
Finally, this information is integrated into the final \ac{BOMI}.

\begin{lstlisting}[caption={Excerpt from the dynamic code index containing a runtime generated class in the \ac{BOMI}.}, label=lst:dynamic-code-index,language=JSON,belowskip=\baselineskip]
  {"com/sun/proxy/$Proxy14":[{"version":"49.0","checksum":"2807adf"}]}
  {"jdk/internal/reflect/GeneratedConstructorAccessor5":[{"checksum":"c636864"}]}
  {"java/lang/invoke/BoundMethodHandle$Species_LLL":[{"checksum":"4f886a8"}]}
  {"jdk/nashorn/internal/scripts/Script$\^eval\_":[{"checksum":"6523d19"}]}
  {"io/dropwizard/jersey/DropwizardResourceConfig$SpecificBinder5e093dc6-5884-44cc-9901-1417d447e561":[{"hash":"41f6c89"}]}
  ...
\end{lstlisting}

The dynamic code generation indexer is run on projects verified with application level integrity checks: the study subjects provide release checksum and signed releases so that they can be verified.

\subsection{SBOM Runtime Watchdog}
\toolname is meant to detect unknown code before it is being executed in production, in order to fail-fast under attack, see \autoref{sec:threat-model}.
The SBOM Runtime Watchdog is responsible for terminating the application when this happens.
It is a Java agent that is attached to the application during startup and takes as input the \ac{BOMI} generated by the above three indexers.
At runtime, it intercepts all classes that are loaded by the \ac{JVM}, computes the checksum of the classes and verifies if the class is in the \ac{BOMI}. 
If the class exists in the \ac{BOMI} and the checksum matches, it means that the class is known and the application continues to execute.
If the class is unknown, it means that either the application is under attack or that the indexing was incomplete. 
Our systematic analysis of the Java software stack is made to exclude the latter (see our experiments in \autoref{sec:rq-applicability}).

The core duty of the SBOM Runtime Watchdog is to disallow unknown classes and terminate the application when an unknown class is detected.
In addition, the incident is reported to the security team.

\subsection{Bytecode Canonicalization}
\label{sec:bytecode-canonicalization}

A major problem with runtime generated code is that it is non-deterministic.
This means that running the same application twice would give slightly different generated bytecodes. To the best of our knowledge, the Java standard library cannot be forced to be fully deterministic~\cite{noauthor_class_nodate}.
It is capable of producing non-deterministic code~\cite{xiong_build_2022}.
For example, \autoref{lst:proxy-difference} shows the difference between two decompiled versions of the same proxy class in different executions.
To account for that problem, \toolname preprocesses all generated classes to obtain a deterministic canonical version,  which we describe below.

The first addressed non-determinism feature is the class name.
The names of generated classes, like \texttt{Proxy} classes include a number that depends on the order of the class loaded, this order is non-deterministic due to e.g. parallelism.
Hence this number could be different between runs even though the content of the bytecode is equivalent.
In order to ensure that these classes are not considered as different classes, we rewrite the class names to a fixed string constant.
For example, the class \texttt{\$Proxy21} and \texttt{\$Proxy14} in line number 1 and 2 in \autoref{lst:proxy-difference} is rewritten to \texttt{foo} in line number 1 in \autoref{lst:canonicalized-bytecode}.

The second non-addressed non-determinism feature is type references in the class code.
The names of type references are derived from the class name and are also non-deterministic.
For example, if class \texttt{\$Proxy21} has a field called \texttt{m3}.
It can be referred to as \texttt{\$Proxy21.m3}, where \texttt{\$Proxy21} is the type reference.
This reference can be used in the bytecode of the same class as shown in line number 12 of \autoref{lst:proxy-difference} or a different class where \texttt{\$Proxy21} is imported.
Thus, we also need to rewrite these references to a fixed string constant \texttt{foo} like in line number 6 of \autoref{lst:canonicalized-bytecode}.

Finally, a third problem lies in the mapping of fields and methods in generated classes between different runs.
For example, \texttt{m3} is mapped to \texttt{visualUpdate} in one run and \texttt{expert} in another run.
This also affected the order of statements in the methods as shown in line number 30 to 33 of \autoref{lst:proxy-difference}.
This is fixed by sorting the byte array of the bytecode just before computing the checksum in the next step.

After performing the above three transformations, we have a canonical, stable, and deterministic view of the class. \toolname then computes the SHA-256 checksum of the constant pool and the JVM opcodes.
The constant pool of a bytecode reflects all the APIs and their exact arguments.
The JVM opcodes reflect how these APIs and the arguments are combined and executed.

To sum up, bytecode canonicalization is an essential component of \toolname. Without it, it is impossible to reason about the integrity of binary code across executions. To our knowledge, we are the first to propose such a canonicalization of runtime generated code for the JVM.

\begin{lstlisting}[backgroundcolor=\color{gray!10}, style=diff,caption={Difference between decompiled output of the same proxy class based on java.beans.BeanProperty, showing non-deterministic naming conventions.}, label=lst:proxy-difference,language=Java,belowskip=\baselineskip]
%\RHilight%-public final class $Proxy21 extends Proxy implements BeanProperty {
%\GHilight%+public final class $Proxy14 extends Proxy implements BeanProperty {
      // [Truncated field declarations]  
%\RHilight%-    public $Proxy21(InvocationHandler var1) {
%\GHilight%+    public $Proxy14(InvocationHandler var1) {
          super(var1);
      }
  
%\RHilight%-    public final boolean visualUpdate() {
%\GHilight%+    public final boolean expert() {
          try {
%\RHilight%-            return (Boolean)super.h.invoke(this, $Proxy21.m3, (Object[])null);
%\GHilight%+            return (Boolean)super.h.invoke(this, $Proxy14.m3, (Object[])null);
          } catch (RuntimeException | Error var2) {
              throw var2;
          } catch (Throwable var3) {
          }
      }
  
%\RHilight%-    public final boolean expert() {
%\GHilight%+    public final boolean visualUpdate() {
          try {
%\RHilight%-            return (Boolean)super.h.invoke(this, $Proxy21.m4, (Object[])null);
%\GHilight%+            return (Boolean)super.h.invoke(this, $Proxy14.m4, (Object[])null);
          } catch (RuntimeException | Error var2) {
              throw var2;
          } catch (Throwable var3) {
  
          try {
%\RHilight%-            $Proxy21.m3 = Class.forName("java.beans.BeanProperty", false, var0).getMethod("visualUpdate");
%\GHilight%+            $Proxy14.m3 = Class.forName("java.beans.BeanProperty", false, var0).getMethod("expert");
%\RHilight%-            $Proxy21.m4 = Class.forName("java.beans.BeanProperty", false, var0).getMethod("expert");
%\GHilight%+            $Proxy14.m4 = Class.forName("java.beans.BeanProperty", false, var0).getMethod("visualUpdate");
          } catch (NoSuchMethodException var2) {
              throw new NoSuchMethodError(var2.getMessage());
          } catch (ClassNotFoundException var3) {
\end{lstlisting}

\begin{lstlisting}[backgroundcolor=\color{gray!10}, caption={\toolname canonicalization allows for stable checksums in the BOMI.}, label=lst:canonicalized-bytecode,language=Java,belowskip=\baselineskip]
  public final class foo extends Proxy implements BeanProperty {
    // [Truncated field declarations]  

    public final boolean visualUpdate() {
        try {
            return (Boolean)super.h.invoke(this, foo.m3, (Object[])null);
        } catch (RuntimeException | Error var2) {
            throw var2;
        } catch (Throwable var3) {
            throw new UndeclaredThrowableException(var3);
        }
    }

    public final boolean expert() {
        try {
            return (Boolean)super.h.invoke(this, foo.m4, (Object[])null);
        } catch (RuntimeException | Error var2) {
            throw var2;
        } catch (Throwable var3) {
            throw new UndeclaredThrowableException(var3);
        }
    }

    // [Truncated methods]
}
\end{lstlisting}

\subsection{BOMI Integrity}
The \ac{BOMI} is a critical component of \toolname.
We take the following special measures to ensure the integrity of the BOMI.
We ensure that all the indexing processes are running in a trusted environment.
We create a Docker image from the base image of the official OpenJDK~\cite{noauthor_openjdk_nodate} and then install  \toolname in the container.

First, the environment indexer should analyze a safe version of the Java standard library.
This is ensured by our selection of the official Java distribution.

Second, we ensure that the indexers interact with trusted sources.
For the supply chain index, we only consider the classes that are downloaded from Maven Central. The Maven Central repository has many sub repositories, we ensure that the URLs of the JAR belong to one of the sub repositories of Maven Central.
We also install the SSL certificates required to interact with Maven Central so that dependency resolution uses verified URLs.
Enforcing SSL certificate verification also prevents DNS hijacking attacks.

\subsection{Implementation}

We have implemented a prototype of the design which is made available on GitHub~\cite{sharma_sbom_2024} under the same name \toolname.

\section{Experimental Methodology}

To evaluate \toolname, we run experiments with 6 study subjects in order to answer the following research questions:

\begin{itemize}
    \item \rqscale (Scale)
    \item \rqeffectiveness (Effectiveness)
    \item \rqapplicability (Applicability)
    \item \rqperformance (Performance)
\end{itemize}

\subsection{Study Subjects}

\begin{table*}[t]
  \centering
  \caption{Study subjects used in the evaluation of \toolname.}
  \label{tab:study-subjects}
  \begin{tabular}{|l|l|l|l|}
  \hline
  \multicolumn{1}{|c|}{\textbf{Study Subjects}} & \multicolumn{1}{c|}{\textbf{Java Version}} & \multicolumn{1}{c|}{\textbf{Dependencies}} & \multicolumn{1}{c|}{\textbf{Workload}} \\ \hline
  {\href{https://nvd.nist.gov/vuln/detail/CVE-2021-44228}{\texttt{Log4Shell} - CVE-2021-44228}\cite{noauthor_nvd_nodate}}          & 17.0.10
  Temurin         &  2        &    String input for logging   \\ \hline
  \href{https://nvd.nist.gov/vuln/detail/CVE-2021-42392}{CVE-2021-42392}\cite{noauthor_nvd_nodatea}                                 &         17.0.10
  Temurin                       &     1                           &      Authentication with server                          \\ \hline
  \href{https://nvd.nist.gov/vuln/detail/CVE-2022-33980}{CVE-2022-33980}\cite{noauthor_nvd_nodateb}                                 &   11 GA OpenJDK                             &    5                            &             JS Script to print date                   \\ \hline
  \href{https://github.com/apache/pdfbox}{apache/pdfbox}\cite{noauthor_apache_nodate}                                &     21.0.2 OpenJDK                           &        12                        &         10 PDF manipulation commands                       \\ \hline
  \href{https://github.com/mpetazzoni/ttorrent}{mpetazzoni/ttorrent}\cite{petazzoni_mpetazzoni_2024}                                &            21.0.2 OpenJDK                    &        6                        &    Torrent downloading                            \\ \hline
  \href{https://github.com/graphhopper/graphhopper/}{graphhopper/graphhopper}\cite{noauthor_graphhopper_2017}                                &         21.0.2 OpenJDK                       &          165                      &      Server initialization and 5 routing requests                          \\ \hline
  \end{tabular}
\end{table*}

We document the study subjects in \autoref{tab:study-subjects} and these will be used to evaluate \toolname.
The first three subjects are high-profile \ac{CVE} with severity rating critical associated with them.
They fall under our threat model and are used to evaluate the effectiveness of \toolname.
The last three subjects are real-world applications that are used to evaluate the applicability and performance of \toolname.
These Java applications are open-source and popular repositories on GitHub.
The next column shows the Java version used to run the application.
The versions are the latest LTS, but for CVE, older LTS versions are used to replicate the vulnerability.
Next, we show the number of dependencies that are part of the application.
Finally, we show the workload that is used for three different purposes.
1) For replicating the vulnerability in the case of high-profile attacks.
2) For running the study subject under the influence of \ac{SBOM} Runtime Watchdog to evaluate the applicability of \toolname.
And finally, 3) for measuring the overhead of \toolname.

\subsection{RQ1: Scale of \ac{BOMI}}

To answer this question, we study \ac{BOMI}s generated by \toolname for the 3 CVEs and 3 real-world applications that will be studied in RQ2 and RQ3.
We create the \bomienvironment statically using the Java Standard Library.
We select \texttt{build-info-go} as the \ac{SBOM} producer to generate the \bomisupplychain.
It is deemed to be the most precise way to capture the dependencies for a Java application~\cite{balliu_challenges_2023}.
Finally, test suites are used by Dynamic Code Indexer to create the \bomiruntime.
Then, we manually analyze them and comment on their characteristics.
We also, comment on how deterministic the generation of \ac{BOMI} is.
The \ac{BOMI} is deterministic if the same classes and checksums corresponding to them are generated for the same application across different runs.

\subsection{RQ2: Effectiveness of \toolname}

For RQ1, we collect three high-profile attacks to show that \toolname can successfully stop them.
High-profile attacks are those that have a \ac{CVE} with a severity rating critical associated with them.
We first replicate the vulnerability in a proof-of-concept application and we show how it can be exploited using a malicious payload.
Next, we create a \ac{BOMI} for the application and pass it to \toolname for enforcement at runtime.
Finally, we deploy the application with \toolname's \ac{SBOM} Runtime Watchdog and we run the attack using the same malicious payload as in the initial reproduction.
We check that \toolname's approach works and that the application terminates before the malicious class is executed, completely mitigating the attack.

\subsection{RQ3: Applicability}

We ensure that \toolname can be used for providing runtime integrity to real-world applications without any false positives, i.e. without terminating the application wrongfully.
To answer this question, we run \toolname on a set of 3 real-world applications - \texttt{PDFBox} (an application for manipulating PDF files), \texttt{GraphHopper} (an open source navigation engine that powers \href{openstreetmap.org}{openstreetmap.org}) \texttt{Ttorrent} (a peer-to-peer
file downloading tool based on the BitTorrent protocol).
Then we construct the \ac{BOMI} for each application.
Finally, we run the application with the workload, while attaching the Runtime SBOM Watchdog to the \ac{JVM}.
If the application executes successfully without inappropriate termination by the watchdog (false positive), we consider the application to be compatible with \toolname.


\subsection{RQ4: Overhead of \toolname}

We measure the overhead of \toolname caused by the \ac{SBOM} Runtime Watchdog by running all study subjects with appropriate workloads.
The workload is the same as used in RQ3 to evaluate the applicability of \toolname.
These workloads are also appropriate for measuring the overhead of \toolname as they are representative of the real-world usage of the application.
Per the best practices of overhead measurement in Java, the evaluation is done using a microbenchmarking framework \ac{JMH} ~\cite{noauthor_openjdk_nodatea} and it reports two metrics - end-to-end time with warmup and workload time excluding warmup.
End-to-end time with warmup is the measure of the sum of how long it takes for the \ac{JVM} to warm up and then run the application.
Warming up of \ac{JVM} is the process of profiling and compiling the bytecode to machine code, it is significant because the JVM has two JIT compilers.
We configure \ac{JMH} to spawn 5 forks of \ac{JVM} wherein it executes 5 warm-up executions first so that the just-in-time compilers can perform optimizations and compile bytecode to machine code.
After that, it runs 5 measurement executions of the workload.
End-to-end time with warmup reports the average of 5 warm-up runs and 5 measurement runs over 5 forks.
Workload time excluding warmup reports the mean of the runtime of the application over all executions and it excludes the \ac{JVM} warm-up time.
Hence, it gives the execution time when all code has been compiled to machine code and the application is running in its most optimized state.
So, times excluding warmup are always lower than end-to-end times with warmup.
Finally, we report the percentage overhead for the two metrics.
We use the following formula to calculate the overhead:
\begin{gather*}
  Actual\ Value = \frac{t_{\text{w/}} - t_{\text{w/o}}}{t_{\text{w/o}}}
  \\
  Error = \frac{t_{\text{w/}} - t_{\text{w/o}}}{t_{\text{w/o}}}\sqrt{\left(\frac{\Delta t_{\text{w/}} + \Delta t_{\text{w/o}}}{t_{\text{w/}} - t_{\text{w/o}}}\right)^2 \pm \left(\frac{\Delta t_{\text{w/o}}}{t_{\text{w/o}}}\right)^2}
  \\
  Overhead = \left(Actual\ Value \pm Error\right) \times 100 \%
\end{gather*}
where
\begin{itemize}
  \item $t_{\text{w/}}$ is the time reported with \toolname,
  \item $t_{\text{w/o}}$ is the workload time without \toolname,
  \item $\Delta t_{\text{w/}}$ is the error in time with $t_{\text{w/}}$,
  \item $\Delta t_{\text{w/o}}$ is the error in time without $t_{\text{w/o}}$,
  \item $Actual\ Value$ is the relative overhead incurred by \toolname,
  \item $Error$ is the error in the overhead calculation.
\end{itemize}

\section{Experimental Results}

We present the results for all the research questions in the following subsections.
All the experiments are run on Azure's virtual machine called Standard D2as v4.
It runs Ubuntu 22.04 LTS (64-bit).
The underlying hardware consists of 8GB RAM and 2 virtual cores of Intel® Xeon® Platinum 8272CL.
We also host all the results on the GitHub repository~\cite{sharma_sbom_2024}.

\subsection{RQ1: Scale of BOMI}
\label{sec:rq-robustness}

\begin{table*}[ht]
  \centering
  \caption{\ac{BOMI} for all the study subjects.}
  \label{tab:bomi-scale}
  \begin{tabular}{|l|l|l|l|l|l|}
  \hline
  \multicolumn{1}{|c|}{\textbf{Study Subjects}} & \multicolumn{1}{c|}{\textbf{\bomienvironment}} & \multicolumn{1}{c|}{\textbf{\bomisupplychain}} & \multicolumn{1}{c|}{\textbf{\bomiruntime}} & \multicolumn{1}{c|}{\textbf{\ac{BOMI}}} & \multicolumn{1}{c|}{\textbf{Reproducible}} \\ \hline
  \href{https://nvd.nist.gov/vuln/detail/CVE-2021-44228}{CVE-2021-44228}      & 24085 & 1268  & 27  &       25381                         &               \cmark                 \\ \hline
  \href{https://nvd.nist.gov/vuln/detail/CVE-2021-42392}{CVE-2021-42392}                          &     24085                  &        944               &          20             &          25049                      &          \cmark                      \\ \hline
  \href{https://nvd.nist.gov/vuln/detail/CVE-2022-33980}{CVE-2022-33980}                        &    24278                   &       788                &            34           &          25100                      &         \cmark                       \\ \hline
  \href{https://github.com/apache/pdfbox}{apache/pdfbox}                        &        25309               &         17760              &             83       &            43152                    &            \cmark                    \\ \hline
  \href{https://github.com/mpetazzoni/ttorrent}{mpetazzoni/ttorrent}                        &       25309                &       801                &      9                 &            26119                    &              \cmark                  \\ \hline
  \href{https://github.com/graphhopper/graphhopper/}{graphhopper/graphhopper}                        &      25309                 &  23685                     &    266                   &            49260                    &             \xmark                   \\ \hline
  \end{tabular}
\end{table*}

\autoref{tab:bomi-scale} presents the details of the \ac{BOMI} for all the study subjects.
The first column lists the study subjects that will evaluate the effectiveness and applicability of \toolname in RQ2 and RQ3 respectively.
The second column is the number of classes in \bomienvironment.
It is dependent upon the Java version.
The third column is the number of classes in \bomisupplychain.
The classes in the supply chain include the dependencies and the project itself.
The fourth column lists the number of classes in \bomiruntime.
\ac{BOMI} shows the total number of classes by adding the classes from \bomienvironment, \bomisupplychain, and \bomiruntime.
Finally, the last column reports if the \ac{BOMI} is reproducible or not.
The \ac{BOMI} is reproducible if all classes and checksums in \bomienvironment, \bomisupplychain, and \bomiruntime are deterministic.

For CVE-2021-44228, we build the \bomienvironment using the Java standard library version 17.0.10 Temurin and we capture 24085 classes.
The \bomisupplychain contains 1269 classes from the source code for replication on this CVE and its 2 dependencies.
The \bomiruntime contains 27 classes that are generated during the execution of the test suite of replication.
The total number of classes in the \ac{BOMI} is 25381.
We generated the \ac{BOMI} twice and it contained the same classes and checksums.
Hence, we consider the \ac{BOMI} for CVE-2021-44228 to be reproducible.

We discuss characteristics of \bomienvironment, \bomisupplychain, and \bomiruntime in the following subsections.

\subsubsection*{\bomienvironment}

The \bomienvironment contains classes from the Java standard library.
Its size is solely dependent upon two factors - the exact version and the vendor of the Java standard library.
In our experiments, the classes in \bomienvironment contributed significantly to \ac{BOMI} no matter the size of the application.

\subsubsection*{\bomisupplychain}

The \bomisupplychain varies significantly across the study subjects as it depends on the number of dependencies and the size of the project.
The size of classes in \bomisupplychain in the \ac{CVE}s is smaller compared to the real-world applications because we replicate the vulnerability in a small proof-of-concept application.
For example, we only need to include one library, \href{https://central.sonatype.com/artifact/com.h2database/h2/1.4.200}{com.h2database:h2:1.4.200}, and 1 class to replicate the vulnerability in CVE-2021-42392.
However, the number of classes in \bomisupplychain is not directly determined by the number of dependencies.
\texttt{PDFBox} has 12 dependencies and the number of classes in \bomisupplychain is 17760.
\texttt{Ttorrent} has half the number of dependencies compared to \texttt{PDFBox} but the number of classes is 20 times less.
This implies that the supply chain of \texttt{Ttorrent} contains smaller dependencies compared to \texttt{PDFBox}.
Finally, the number of classes in \bomisupplychain of \texttt{GraphHopper} is comparable to the classes in \bomiruntime, but they have very different characteristics.
Classes in \bomiruntime are provided by the Java standard library.
All those classes are maintained by the same organization.
It is consumed by many more stakeholders and each class is under high scrutiny for quality by developers of Java standard library.
However, classes in \bomisupplychain are provided by different organizations or open-source developers and are developed independently of each other.
Moreover, the classes in \bomisupplychain come from a huge dependency tree containing direct and indirect dependencies.
For example, developers of \texttt{GraphHopper} declare 4 dependencies and those dependencies have 161 dependencies.
The relationships between these 165 dependencies result in the tree being 5 levels deep where only the first level dependencies are declared by the developers.
This makes the classes in \bomisupplychain very diverse compared to the classes in \bomiruntime.

\subsubsection*{\bomiruntime}

\bomiruntime is much smaller compared to the last two indices, but these classes are extremely important to capture.
Our threat model targets the attacks that are possible because dynamic classes are executed.
These classes are not present in Java application JARs nor in the \ac{SBOM} and they only appear at runtime.
A proof-of-concept application for CVE-2021-44228 has 27 classes in \bomiruntime.
If we enforce that only those classes are allowed to run, we can prevent the loading of malicious classes and attack can be mitigated.

\subsubsection*{Reproducibility}

Reproducibility is an important property for \ac{BOMI}.
The \ac{BOMI} must list the exact classes and checksums for a specific version of the application.
Otherwise, the enforcement of \ac{BOMI} could cause inconsistent termination of the application.
We observe that \bomienvironment and \bomisupplychain are deterministic for all classes.
However, \bomiruntime is non-reproducible for \texttt{GraphHopper} because one of its dependencies, dropwizard, generates classes based on a random UUID.
These classes are different each time the \bomiruntime is generated and hence the overall \ac{BOMI} for \texttt{GraphHopper} is also not reproducible.
There is also non-determinism in other dynamic classes but is taken care of by our bytecode canonicalization process~\autoref{sec:bytecode-canonicalization}.
The \ac{BOMI} for all the other study subjects is reproducible.

\begin{mdframed}\noindent
  \textbf{Answer to \rqscale} \\
  \ac{BOMI} contains thousands of classes and their checksums.
  The \bomienvironment contributes significantly to the \ac{BOMI} size for all the study subjects due to the sheer complexity of the JVM Standard Library.
  The size of the \bomisupplychain varies significantly across the study subjects depending on the application supply chain.
  Finally, the \bomiruntime is much smaller compared to the other two, but, as we demonstrate later, the classes captured there are essential to prevent attacks due to the execution of dynamic classes at runtime while remaining compatible with existing applications.
\end{mdframed}

\subsection{RQ2: \toolname Effectiveness}
\label{sec:rq-effectiveness}

\begin{table*}[t]
  \centering
  \caption{\ac{CVE}s proven to be fully mitigated by \toolname.}
  \label{tab:cve-mitigated}
  \begin{tabular}{|l|p{0.2\linewidth}|p{0.15\linewidth}|p{0.2\linewidth}|p{0.2\linewidth}|}
  \hline
  \multicolumn{1}{|c|}{\textbf{CVE}} & \multicolumn{1}{c|}{\textbf{Vulnerable Dependency}} & \multicolumn{1}{c|}{\textbf{Use Case}} & \multicolumn{1}{c|}{\textbf{Vulnerable API}} & \multicolumn{1}{c|}{\textbf{Malicious Class}} \\ \hline
  \href{https://nvd.nist.gov/vuln/detail/CVE-2021-44228}{CVE-2021-44228 (\texttt{Log4Shell})}                                &  \href{https://central.sonatype.com/artifact/org.apache.logging.log4j/log4j-core/2.14.1}{org.apache.logging.log4j:log4j-core:2.14.1}      &  Logging framework                 &  \texttt{org.apache.logging.log4j.Logger\#error(String)}  &     \texttt{xExportObject}                           \\ \hline
  \href{https://nvd.nist.gov/vuln/detail/CVE-2021-42392}{CVE-2021-42392}                                &  \href{https://central.sonatype.com/artifact/com.h2database/h2/1.4.200}{com.h2database:h2:1.4.200}        &     Database engine                  & \texttt{org.h2.util.JdbcUtils\#getConnection (String, String,Properties)}    & \texttt{xExportObject}                            \\ \hline
  \href{https://nvd.nist.gov/vuln/detail/CVE-2022-33980}{CVE-2022-33980}                                &  \href{https://central.sonatype.com/artifact/org.apache.commons/commons-configuration2/2.7}{org.apache.commons:commons-configuration2:2.7}    & Parsing configuration files      &      \texttt{org.apache.commons.configuration2.interpol.Lookup\#lookup(String)}              &  \texttt{jdk.nashorn.internal.scripts.Script\$\string\^eval\string\_}                             \\ \hline
  \end{tabular}
\end{table*}

\autoref{tab:cve-mitigated} presents the attacks we aim to mitigate in our experiment.
We first mention the \ac{CVE}s, then we report the vulnerable dependency that includes the \ac{CVE}, as deployed on Maven Central.
The dependency name is the concatenation of the group ID, artefact ID, and version.
We also mention the use case for the dependency, to give some context about where the vulnerability is exploited.
Next, we mention the particular API entrypoint for the attack and thus, make the dependency vulnerable.
Finally, the column `Malicious Class' is the classname that is injected during the attack, and which should be detected and blocked by \toolname.
All the attacks are available in our repository~\cite{sharma_sbom_2024} and are fully reproducible.

For example, consider the first row of the table which has details about \ac{CVE}-2021-44228, also known as \texttt{Log4Shell}.
The vulnerable dependency group ID, artefact ID, and version are \texttt{org.apache.logging.log4j}, \texttt{log4j-core}, and \texttt{2.14.1} respectively.
This dependency is one of the most popular logging frameworks in Java~\cite{cheng_platform_2018}.
The API that processes the malicious input is \texttt{org.apache.logging.log4j.Logger\#error(String)}.
and  the malicious class we use in our experiment is called \texttt{xExportObject}.

\subsubsection*{CVE-2021-44228 (Log4Shell)}

\paragraph*{Replication}
\label{sec:log4j-replication}

We first replicate the \texttt{Log4Shell} vulnerability in a proof-of-concept application.
The application has a single class that has a method containing the vulnerable API whose normal usage is to log messages.
Next, we setup the server hosting a classfile that contains bash commands as explained in \autoref{sec:threat-model}.
We call this classfile \texttt{xExportObject}.
To trigger remote class loading, we pass \texttt{\$\{jndi:ldap://localhost:1389/o=reference\}} as the malicious input to the application.
This looks up the LDAP entry corresponding to the name ``\texttt{o=reference}''.
The entry has attributes, \texttt{javaCodebase} and \texttt{javaFactory}~\cite{seligman_schema_1999}, that are used to load the malicious class \texttt{xExportObject}.
The name of the malicious class is arbitrary and can be any name.
Finally, the malicious class executes the bash command and compromises the application.

\paragraph*{Mitigation}

The mitigation process in \toolname is a two phase process as shown in \autoref{fig:overview}.
The first phase happens offline, and \toolname creates the \ac{BOMI} with the classes from the Java standard library, the supply chain classes and the dynamically generated classes.
For this example, the \bomienvironment gathers all the 24085 classes from the Java standard library version 17.0.10+7 Temurin.
The \bomisupplychain contains the classes from the supply chain of the example application, which enumerates to 1269.
For building the \bomiruntime, we execute a minimal test suite in order to capture dynamic classes.
After running the tests, the dynamic code indexer reports 27 classes.
All of these classes are either Proxy classes or subclasses of \texttt{MagicAccessorImpl} used for introspection (see \autoref{sec:dynamicity-java}).
One of the proxy class implements the interface \texttt{org.apache.logging.log4j.core.config.plugins.Plugin}.
Recall that the \ac{BOMI} needs to contain all the non-malicious runtime classes otherwise our agent can terminate the process over a benign class that was not recorded in the \bomiruntime.

The second phase of the mitigation procedure of \toolname happens at runtime.
In this part of the experiment, we run the application with the same malicious input as before \${jndi:ldap://localhost:1389/o=reference}, and we attach the \ac{SBOM} Runtime Watchdog to the application.
The \ac{JVM} loads classes but it is terminated just before loading the malicious class \texttt{xExportObject}, demonstrating the success of \toolname in mitigating the attack.

We explain this in more detail using \autoref{lst:jndi-call-stack}.
It begins with the (1) main method of the application that attempts to (2) log a message, a malicious input in this case.
The log message is processed by the internal \texttt{Log4j} classes that invoke the (3) lookup method inside the \texttt{Log4j} dependency.
This lookup method is delegated to the (4) \texttt{LdapCtx} class which is part of the Java standard library and it queries for the entry \texttt{o=reference} in the LDAP server.
This query triggers class loading of the malicious class \texttt{xExportObject} by invoking (5) \texttt{DirectoryManager.getObjectInstance()}.
The method reads two attributes of the LDAP entry \texttt{o=reference} - \texttt{javaCodebase} and \texttt{javaFactory}, then (6) \texttt{VersionHelper.loadClass()} initiates class loading~\cite{seligman_schema_1999}.
At this point, the class is integrated into the runtime by the (7) Classloader and is ready for execution.
However, the \ac{SBOM} Runtime Watchdog (8) intercepts the malicious class, \texttt{xExportObject} before execution and terminates the application.
This demonstrates that \toolname can integrate into the complex workflow of modern Java applications and mitigate this high-profile real-world attack.

\begin{lstlisting}[backgroundcolor=\color{gray!10}, style=customCallStack, caption={Call stack at the time when the vulnerable application using Log4j is terminated.}, label=lst:jndi-call-stack]
  ~SBOM_EXE^.^isLoadedClassAllowlisted$()    (8)
  ~ClassLoader^.^defineClass1$()             (7)
  // [...] internal java classes
  ~VersionHelper^.^loadClass$()              (6)
  ~DirectoryManager^.^getObjectInstance$()   (5)
  ~LdapCtx^.^c_lookup$()                     (4)
  ~JndiLookup^.^lookup$()                    (3)
  // [...] internal Log4j classes
  ~Logger^.^log$()                           (2)
  ~Main^.^main$()                            (1)
\end{lstlisting}

\subsubsection*{CVE-2021-42392 (H2)}

\paragraph*{Replication}
\label{sec:h2-replication}

\begin{figure}
  \centering
  \includegraphics[width=\columnwidth]{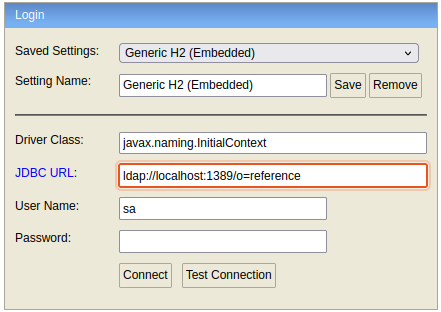}
  \caption{Login screen of the H2 database engine with the malicious input and the default user credentials.}
  \label{fig:h2-database-engine}
\end{figure}

This CVE corresponds to the \texttt{H2} database engine.
To replicate an attack, we create the main class of an application that  starts the database engine indefinitely.
The default page is a login form that requires authentication as shown in \autoref{fig:h2-database-engine}.
We enter the malicious input in the fields "Driver Class" and "JDBC URL" with values \texttt{javax.naming.InitialContext} and \texttt{ldap://localhost:1389/o=reference} respectively.
This leads to remote class loading of the malicious class \texttt{xExportObject} from the same LDAP server as in the previous attack.

\paragraph*{Mitigation}

First, we generate the \ac{BOMI} for this application.
The \bomisupplychain for \texttt{com.h2database:h2:1.4.200} contains 944 classes.
We create a new test to capture the dynamic classes generated during the execution of the application.
The test that starts the database engine, authenticates the default user with username \texttt{sa} and no password as shown in \autoref{fig:h2-database-engine}, and then shuts down the server.
Based on this test, the \bomiruntime records 20 runtime classes that are similar in nature to the ones we found in  the previous application.

In the second phase, we start the server again but with SBOM Runtime Watchdog attached.
Next, we go to the local host and pass the malicious inputs in the fields "Driver Class" and "JDBC URL" with values \texttt{javax.naming.InitialContext} and \texttt{ldap://localhost:1389/o=reference} respectively as shown in \autoref{fig:h2-database-engine}.
The driver class \texttt{javax.naming.InitialContext} triggers the subset of the call stack from (4) to (8) as shown in \autoref{lst:jndi-call-stack}.
Similar to before, the application is terminated before the malicious class \texttt{xExportObject} is executed.
\toolname successfully mitigates CVE-2021-42392.


\subsubsection*{CVE-2022-33980 (Apache Commons)}

\paragraph*{Replication}
\label{sec:apache-commons-replication}

Finally, we consider the CVE-2022-33980 which applies to the \texttt{Apache Commons Configuration} dependency.
This vulnerability leverages the Nashorn JavaScript engine which was bundled with the Java standard library until Java 15.
The proof of concept application has a single class, which invokes the vulnerable API.
The API takes a String input in the form of \texttt{prefix:name}.
For example, \texttt{java} and \texttt{date}~\cite{noauthor_defaultlookups_nodate} can be used to fetch information about the Java runtime and the current date respectively.
We pass the malicious input \texttt{javascript:var clazz = java.lang.invoke.MethodHandles.lookup().defineClass(<malicious bytecode>); m = clazz.getMethod('main'); m.invoke(null);}.
The input triggers the Nashorn engine, which generates the class \texttt{jdk.nashorn.internal.scripts.Script\$\string\^eval\string\_} and executes the malicious bytecode.

\paragraph*{Mitigation}

To experiment with \toolname against this attack, we first create the \ac{BOMI} for the application.
The \bomienvironment contains the 24278 classes of Java 11+28 OpenJDK.
The \bomisupplychain contains 788 classes from the \texttt{org.apache.commons:commons-configuration2:2.7} and its dependencies.
To create the \bomiruntime, we run a test that evaluates a JavaScript code printing date and time to the terminal.
Based on this test, the \bomiruntime contains 34 classes.

\begin{lstlisting}[float,backgroundcolor=\color{gray!10}, style=customCallStack,caption={Call stack at the time when the vulnerable application using commons-configuration is terminated when \toolname detects the malicious class.}, label=lst:nashorn-call-stack]
  ~SBOM_EXE^.^isLoadedClassAllowlisted$()         (7)
  ~ClassLoader^.^defineClass1$()                  (6)
  // [...] internal java classes
  ~CompilationPhase$InstallPhase^.^transform$()   (5)
  ~Compiler^.^compile$()                          (4)
  // [...] internal nashorn classes
  ~NashornScriptEngine^.^eval$()                  (3)
  ~StringLookupAdapter^.^lookup$()                (2)
  ~Main^.^main$()                                 (1)
\end{lstlisting}

We run the application with a malicious JavaScript code as input in the form \texttt{"prefix:name"}.
The input contains a malicious bytecode array and JavaScript APIs that would be loaded and executed.
We explain the mitigation of vulnerability using the call stack in \autoref{lst:nashorn-call-stack}.
(1) Main code of the application takes in the input and (2) the class \texttt{StringLookupAdapter} in \texttt{org.apache.commons:commons-configuration2:2.7} processes it.
Since the prefix is \texttt{javascript}, it triggers the (3) Nashorn engine to evaluate the JavaScript code.
The Nashorn engine (4) compilation has 13 phases.
The bytecode corresponding to the JavaScript code is generated in the 12th phase,`Bytecode Generation', and then in the (5) final phase, the bytecode class loading is initiated.
The class is then integrated into the runtime by the (6) Classloader and is ready for execution.
However, the \ac{SBOM} Runtime Watchdog (7) intercepts the malicious class before execution and terminates the application.
\toolname successfully stops the attacker of CVE-2021-42392.

\begin{mdframed}\noindent
  \textbf{Answer to \rqeffectiveness} \\
  \toolname successfully indexes a comprehensive allowlist for high-profile \ac{CVE}s that are exploitable. At runtime, the \ac{SBOM} Watchdog of \toolname does detect and prevent the execution of malicious classes, because they are not in the \ac{BOMI}, terminating the vulnerable application before it is infected. \toolname mitigates \texttt{Log4Shell}, one of the most devastating software supply chain attacks in years.
\end{mdframed}

\subsection{RQ3: \toolname Applicability}
\label{sec:rq-applicability}

\begin{table*}[t]
  \centering
  \caption{Replications used to evaluate whether \toolname is compatible with real-world code, without spurious terminations.}
  \label{tab:real-world-apps}
  \begin{tabular}{|l|l|l|l|l|l|l|}
  \hline
  \multicolumn{1}{|c|}{\textbf{Project}} & \multicolumn{1}{c|}{\textbf{Module}} & \multicolumn{1}{c|}{\textbf{Use Case}} & \multicolumn{1}{c|}{\textbf{Release}}  & \multicolumn{1}{c|}{\textbf{kLOC}} & \multicolumn{1}{c|}{\textbf{Tests}} & \multicolumn{1}{c|}{\textbf{Stars}} \\ \hline
  \href{https://github.com/apache/pdfbox}{apache/pdfbox}      & \texttt{pdfbox-app} & PDF manipulation   &    \href{https://mvnrepository.com/artifact/org.apache.pdfbox/pdfbox-app/3.0.2}{3.0.2}       &    117  & 1805 & 2.5K  \\ \hline
  \href{https://github.com/mpetazzoni/ttorrent}{mpetazzoni/ttorrent}     & \texttt{ttorrent-cli} & Torrent client    &   \href{https://mvnrepository.com/artifact/com.turn/ttorrent-cli/1.5}{1.5}   & 6 & 4 & 1.4K \\ \hline
  \href{https://github.com/graphhopper/graphhopper/}{graphhopper/graphhopper}      & \texttt{graphhopper-web} &  Navigation engine  &    \href{https://mvnrepository.com/artifact/com.graphhopper/graphhopper-web/9.1}{9.1}     &    56  & 2878 & 4.8K \\ \hline

  \end{tabular}
\end{table*}




We select the versions of the real-world applications as shown in \autoref{tab:real-world-apps} to evaluate the compatibility of \toolname.

The first column `Project'  is the name of the GitHub repository of the application.
The module column is the exact maven module that we consider.
The use case column describes the functionality of the application that we evaluate.
The release column is the latest version of the application as of 1\textsuperscript{st} June 2024, hosted on Maven Central.
kLOC indicates the number of thousands of lines of code in the module.
The number of tests is the total number of tests that are executed to capture the dynamic classes.
Finally, we indicate the number of stars the repository has on GitHub, as an indicator of the popularity of the application.

For example, in the first row, we consider the \texttt{PDFBox} application.
We consider the module \texttt{pdfbox-app} as it is the main module for performing PDF manipulations.
The main module in maven projects produces the executable JAR.
We take the latest release \texttt{3.0.2} of the module.
The application has 117 kLOC and 907 tests.
The repository has 2.5K stars on GitHub.

In the following subsections, we present the results for each application.
We run our experiments using OpenJDK 21.0.2.
We organize the results of each application in four parts.

First, we describe the creation of a workload so that we can prepare a baseline usage for the application, as described in~\autoref{tab:study-subjects}.
Then we describe the generated \ac{BOMI} and we report the classes detected by \toolname that are not part of the \ac{BOMI} and the reasons those classes are not there.
Finally, we discuss the changes that we made to the application to make it compatible with \toolname.

\subsubsection*{Successful run of PDFBox}

The workload for \texttt{PDFBox} includes 10 PDF manipulations that can be done using the \texttt{pdfbox-app} module~\cite{noauthor_apache_nodate}.
They consist of manipulating a sample PDF file and verifying the output.

Before running the workload, we create the \ac{BOMI} for the application.
The \bomienvironment contains 25309 classes from the Java standard library version 21.0.2+13-58 OpenJDK.
The \bomisupplychain contains 17760 classes from all dependencies of \texttt{PDFBox}.
The \bomiruntime contains 83 \texttt{PDFBox} classes that are dynamically loaded during the execution of the test suite.

Next, we run the same workload with the \ac{SBOM} Runtime Watchdog.
We notice that it reports 4 classes that are not allowlisted as shown in \autoref{lst:pdfbox-fp}.
These proxy classes are generated because every manipulation command in PDFBox invokes APIs related to the \texttt{debugger} module of \texttt{PDFBox}.
These APIs are not tested in the existing test suite and hence do not appear in the \ac{BOMI}. This is a consistent, expected behavior.

\begin{lstlisting}[caption={False positives reported by PDFBox.}, label=lst:pdfbox-fp,language=JSON]
  [NOT ALLOWLISTED]: jdk/proxy1/$Proxy11
  [NOT ALLOWLISTED]: jdk/proxy1/$Proxy12
  [NOT ALLOWLISTED]: jdk/proxy1/$Proxy13
  [NOT ALLOWLISTED]: jdk/proxy1/$Proxy14
\end{lstlisting}

To make the application fully compatible with \toolname's approach, we write an additional test case.
The test invokes the PDFBox Debugger as GUI and then shuts it down, this test makes sure that the classes generated by the debugger are captured during Dynamic Code Indexing.
Once again, we run the application with the \ac{SBOM} Runtime Watchdog and it runs to completion because all the classes loaded during normal execution are part of the \ac{BOMI}.
Thus, \toolname is compatible with the \texttt{PDFBox} application.

\subsubsection*{Successful run of Ttorrent}

The workload for the \texttt{Ttorrent} application is to download a torrent file and  verify that the download has completed.
We use a torrent file that has 1 text file in it.

The \bomisupplychain contains 801 classes from the \texttt{com.turn:ttorrent-cli:1.5} and its dependencies.
\bomiruntime consists of 9 classes.

Upon initial run of Ttorrent, the \ac{SBOM} Runtime Watchdog reports 11 classes that are not part of the \ac{BOMI}.
One proxy class is generated because of the torrent download.
The rest of the 10 classes are labelled as modified because Maven Central hosts the dependencies of \texttt{Ttorrent} with an outdated version of Java bytecode.
The JAR that we execute contains a more recent version hence there is a difference in checksums.

We make two necessary changes in order to eliminate these initial false positives.
First, we add a test that downloads the torrent file, and this captures the missing proxy class in \bomiruntime.
Second, we self-host \texttt{ttorrent-cli:1.5} and its dependencies.
This ensures that the classes in \bomisupplychain of \texttt{Ttorrent} are compiled to Java 8 bytecode.
After these changes, we are able to download the  torrent and no false positives are reported. 
This demonstrates that for \toolname to work: 1) tests must trigger the generation of runtime classes and 2) package managers have to be kept up to date.

\subsubsection*{Partially successful run of GraphHopper}

We design the workload for the \texttt{GraphHopper} application that starts up the server and makes 5 routing requests within Berlin, Germany.
They involve multiple paths between source, destination, and intermediate points.
For example, one of the requests is a route from \texttt{(13.40, 52.55)} to \texttt{(13.50, 52.50)}.
The first floating point number in the tuple is the longitude and the second is the latitude.

The \bomienvironment is the exact same as the one in \texttt{PDFBox} because we do not change the Java version.
The \bomisupplychain contains 23685 classes, and the \bomiruntime contains 266 classes.

Before requesting the routes, the server initialization reports 5 classes that are not part of the \ac{BOMI}.
The first 3 are proxy classes and are generated in order to start up the server.
The next 2 are related to the configuration of the server. 


We investigate all classes and find two clear reasons why they are not included in \ac{BOMI}.
First, there is no test that initializes the server.
We add a test that initializes the server and then shuts it down when the server is ready, this captures all proxy classes.


Finally, there is a case where the names of the generated classes are based on a random UUID.
Since we do a lookup based on the class name, we are not able to find these classes in the \ac{BOMI}.
Because of this, \toolname is not fully compatible with \texttt{GraphHopper}.
Note that it does not invalidate the soundness of bytecode canonicalization.
We believe that the whole \ac{JVM} architecture has not been designed with determinism in mind, yielding this kind of integrity problems.
In our current landscape dominated by cybersecurity concerns, this is being taken care of now in other ecosystems.
For example, determinism is part of the architecture in Go ecosystem~\cite{cox_perfectly_2023}.

\begin{mdframed}\noindent
  \textbf{Answer to \rqapplicability} \\
  \toolname is fully compatible with real-world applications like \texttt{PDFBox}, and \texttt{Ttorrent} with minimal changes.
  This validates the two key concepts of indexing and bytecode canonicalization.
  Our experiments also show the importance of appropriate testing for capturing all dynamically generated classes in the \ac{BOMI}. 
\end{mdframed}

\subsection{RQ4: \toolname Overhead}
\label{sec:rq-overhead}

\begin{table*}[t]
  \centering
  \caption{Performance and Overhead Measurement for \toolname.}
  \label{tab:performance}
  \begin{tabular}{|lllllll|}
  \hline
  \multicolumn{1}{|c|}{\multirow{3}{*}{\textbf{Study Subjects}}} & \multicolumn{6}{c|}{\textbf{Performance (seconds / number of execution of workload)}} \\ \cline{2-7} 
  \multicolumn{1}{|c|}{} & \multicolumn{3}{c|}{\textbf{End-to-end time with warmup}}                                                                   & \multicolumn{3}{c|}{\textbf{Workload time excluding warmup}}                                              \\ \cline{2-7}
  \multicolumn{1}{|c|}{} & \multicolumn{1}{l|}{\textbf{W/O \toolname}} & \multicolumn{1}{l|}{\textbf{W/ \toolname}} & \multicolumn{1}{l|}{\textbf{Overhead}} & \multicolumn{1}{l|}{\textbf{W/O \toolname}} & \multicolumn{1}{l|}{\textbf{W/ \toolname}} & \textbf{Overhead} \\ \hline
  \href{https://github.com/apache/pdfbox}{apache/pdfbox} & \multicolumn{1}{|l|}{$5.949 \pm 0.281$}          & \multicolumn{1}{l|}{$7.380 \pm 1.456$}          & \multicolumn{1}{l|}{$(24.1 \pm 29.2)\%$}          & \multicolumn{1}{l|}{$1.822 \pm 0.027$}          & \multicolumn{1}{l|}{$1.840 \pm 0.025$}          &     $(1.0 \pm 2.9)\%$      \\ \hline
  \href{https://github.com/mpetazzoni/ttorrent}{mpetazzoni/ttorrent}    & \multicolumn{1}{|l|}{$1.225 \pm 0.465$}          & \multicolumn{1}{l|}{$1.305 \pm 0.108$}          & \multicolumn{1}{l|}{$(6.5 \pm 46.8)\%$}          & \multicolumn{1}{l|}{$0.952 \pm 0.032$}          & \multicolumn{1}{l|}{$0.954 \pm 0.032$}          &      $(0.2 \pm 6.7)\%$     \\ \hline
  \href{https://github.com/graphhopper/graphhopper/}{graphhopper/graphhopper}     & \multicolumn{1}{|l|}{$0.900 \pm 0.098$}          & \multicolumn{1}{l|}{$1.338 \pm 0.264$}          & \multicolumn{1}{l|}{$(48.7 \pm 40.6)\%$}          & \multicolumn{1}{l|}{$0.056 \pm 0.003$}          & \multicolumn{1}{l|}{$0.056 \pm 0.008$}          &      $(0.0 \pm 19.6)\%$    \\ \hline
  \end{tabular}
\end{table*}

\autoref{tab:performance} shows the performance of the study subjects, as recorded by state-of-the-art tool \ac{JMH}~\cite{noauthor_openjdk_nodatea}.
It is given in seconds per execution of the workload.
The second column and the third column list the end-to-end time with warmup and workload time excluding warmup with and without \toolname, as well as the overhead induced by \ac{SBOM} Runtime Watchdog.

For example, \texttt{PDFBox} takes $5.949 \pm 0.281$ and $7.380 \pm 1.456$ seconds to run the workload without and with \toolname respectively when considering the end-to-end time with warmup.
The overhead is computed as follows $\frac{3.404 - 2.551}{2.551} \pm \frac{3.404 - 2.551}{2.551}\sqrt{\left(\frac{0.393+0.204}{3.404 - 2.551}\right)^2 \pm \left(\frac{0.204}{2.551}\right)^2}$.
This evaluates to $(24.1 \pm 29.2)\%$.
Similarly, times after warmup are $1.822 \pm 0.027$ and $1.840 \pm 0.025$ seconds without and with \toolname.
The overhead incurred is $(1.0 \pm 2.9)\%$.

In the following subsections, we discuss the end-to-end time with warmup and workload time excluding warmup metrics.

\subsubsection*{End-to-end time with warmup}

The end-to-end time with warmup is the time taken by the \ac{JVM} to warm up and execute the workload.
While warming up, classes in the order of tens of thousands~\autoref{tab:bomi-scale} are intercepted by the \ac{SBOM} Runtime Watchdog.
They are verified by comparing their checksums with the \ac{BOMI} reference and then allowed to be executed.
This adds a significant overhead to the end-to-end time with warmup.
For example, more than an overhead of $50.0\%$ can be observed in all three study subjects.
End-to-end time with warmup shows that the overhead is not negligible as a result of 1) bytecode canonicalization, and 2) additional classloading work.
We argue it is acceptable in the context of long running applications like \texttt{GraphHopper}.

\subsubsection*{Workload time excluding warmup}

We notice that the overhead is less than 6.9\% in all the study subjects, which is small.
This is because, once the classes are verified by the \ac{SBOM} Runtime Watchdog, they are stored in the \ac{JVM} cache and are not verified anymore.
Hence, the subsequent runs do not have the overhead of verification.
All cases have negligible overhead and that also is incurred during the startup of the application.
Thus, \toolname is suitable for all study subjects, especially for long running applications.

\subsubsection*{Variability in the measurement}

We notice that errors in overhead computed in \autoref{tab:performance} are significant.
This is because of the measurement variability by \ac{JMH}.
The execution time of the workload is dependent upon a variety of factors like CPU allocation, \ac{JVM} interpretation of bytecode, JIT compilation, profiling code, optimizations, garbage collection, etc.
To minimise fluctuations because of these factors, \ac{JMH} runs the benchmark in multiple forks of the \ac{JVM}.
Also, a benchmark could be optimized by the compiler and its execution time reported could be much less than the actual execution time.
This factor does not affect the overhead because the same benchmark is run with and without \toolname.

\begin{mdframed}\noindent
  \textbf{Answer to \rqperformance} \\
  The overhead introduced by \toolname is negligible after warm-up.
  \toolname can be used in production environments, especially with long-running applications, such as web and enterprise application servers, where the warm up time is a one-time cost which is not a concern.
\end{mdframed}

\section{Related Work}

Four strategies have been proposed to protect against malicious code execution at runtime.
In the following subsections, we discuss them.

\subsection{Permission Managers}
Permission managers allow  developers to define access permissions for the application at varying granularities.
Developers define policies, which specify what resources can be accessed by the application, libraries or system calls.
The permission manager then enforces the policies at runtime.
The integrity of the application is protected, as the sensitive resources can only be accessed per the rules defined in the policies.

Amusuo et al.~\cite{amusuo_preventing_2023} map network, filesystem, and process permissions to each dependency in the policy file.
At runtime, the tool enforces the policy by allowing or denying access to the resources based on which access permissions are defined for the dependency.
The enforcement of permissions is done by adding hooks to methods handling resource access in the Java standard library.
Jamrozik et al.~\cite{jamrozik_mining_2016} map the complete application to the Android permissions for location, microphone, etc.
They get the list of resources accessed by running the application against the test generated using fuzzing.
Only the resources that were accessed during the tests are allowed to be accessed at runtime.
Java  also provides native support for writing permissions using \ac{JSM}, which is a built-in permission manager that allows the developer to define access permissions for each Java class.
\ac{JSM} is deprecated for removal~\cite{noauthor_jep_nodate}.

There is a body of literature on permission management for JavaScript.
Ohm et al.~\cite{ohm_you_2023}, Ferreira et al.~\cite{ferreira_containing_2021}, Ahmadpanah et al.~\cite{ahmadpanah_sandtrap_2021}, and De Groef et al.~\cite{degroef_nodesentry_2014} propose permission managers for Node.js applications.
Vasilakis et al.~\cite{vasilakis_preventing_2021} propose enforcing read, write, and execute permissions on each field and method in JavaScript.
Finally, one can enforce permissions on every single system call invoked by an application, such as in SWAT4J~\cite{xu_swat4j_2024}, HODOR~\cite{wang_hodor_2023} and Confine~\cite{rostamipoor_confine_2023}, which leverage Seccomp BPF to restrict invocation of system calls at runtime.

Our approach is fundamentally different because it does not require the developer to define access permissions for the application, libraries, or system calls.
To that extent, our approach is more lightweight and more smoothly applicable in practice.
Moreover, enforcing permissions requires modification in the runtime environment like \ac{JVM} or Node.js which is not required in our approach.

\subsection{Compartmentalization}
Compartmentalization is a method where different parts of an application are executed in different protection domains.
This ensures integrity as one part of the application can be protected against malicious behavior in other parts.
Application partitioning and third-party code isolation are methods that have been studied to compartmentalize applications at runtime.

Application partitioning can be done at the hardware level or the software level.
\ac{SGX} is a hardware-based security feature that runs untrusted code in an isolated environment, called enclave.
Uranus~\cite{jiang_uranus_2020} and Montsalvat~\cite{yuhala_montsalvat_2021} are two frameworks that enable developers to use \ac{SGX} capabilities in their applications.
The developers can use these libraries to indicate trusted and untrusted parts of the application.
The \ac{JVM} will then execute the trusted and untrusted parts in different regions of the CPU.
Another hardware-based mechanism is proposed by Tsampas et al.~\cite{tsampas_automatic_2017}.
They propose a source to source compiler that compiles the code in such a way that it is executed in isolated memory segments.
The isolation of memory segments is done by unforgeable pointers that are protected by hardware.

At software level, the trusted and untrusted part of the application is executed in separate processes.
Another line of work by Song et al.~\cite{song_exploiting_2015} and Gudka et al.~\cite{gudka_clean_2015} relocates the untrusted part of the application to a separate process.
This process has limited access permissions, in the Linux sense.

Isolation of third-party code is another method to compartmentalize applications.
\cite{narayan_retrofitting_2020}, \cite{vasilakis_breakapp_2018}, \cite{lamowski_sandcrust_2017}, and \cite{gama_selfhealing_2010} consider third-party code as untrusted because the developer cannot control its code.
They propose running the third-party code in an isolated environment which has limited access to the system resources.
Then it becomes a challenge for the developer to define the access permissions for the third-party code as we saw in the previous section - access permissions for dependency requires modification in runtimes like Node.js or \ac{JVM}.

\toolname runs the application in a single process so that the performance overhead at runtime is minimal.
Moreover, compartmentalization approaches require manual work to split the application into trusted and untrusted parts.

\subsection{Integrity Measurement}

Integrity Measurement comes from the Linux kernel and means  measuring the application in terms of its call, memory or any kind of execution behavior.
This measurement data is then used at runtime to ensure that the application is executing as expected.
These measurement based techniques can further be divided into two subsections based on the enforcement mechanism.

\subsubsection*{Manual verification of integrity}
These approaches require the user to verify the integrity of the application by analyzing the measurement list.
Hence, the verification of the measurement list is manual.

RIM4J~\cite{ba_rim4j_2018} and JMF~\cite{thober_jmf_2012} propose a remote attestation protocol that verifies the integrity of applications running on the cloud.
Upon request by the client, the cloud application sends the hash of classes to be written in the heap memory to the client.
The client then verifies whether an observed hash is expected and stops the process if the measurement list is tampered.

On a similar line of work, Benedicts et al.~\cite{debenedictis_integrity_2019} and Nauman et al.~\cite{nauman_kernellevel_2010} propose remote attestation protocol for Docker containers and Android platforms respectively.
They leverage techniques similar to \ac{IMA} to measure all the loaded executables.
This measurement list is hosted on a trusted platform and can be used for verification of loaded executables.

\subsubsection*{Automatic verification of integrity}
These approaches attach the measurement list to the process running.
At runtime, the execution behavior is verified automatically based on the measurement list.

RSDS~\cite{wang_rsds_2020} and Prof-gen~\cite{kim_profgen_2021} are two approaches that create an allowlist of system calls that can be invoked by containers.
They employ static and dynamic analysis to get all the invoked system calls and then merge the results to get the final allowlist.
This allowlist is then converted to a seccomp profile and enforced at runtime.

Nomura et al.~\cite{nomura_automatic_2019} propose a method to generate an allowlist of SQL queries that can be invoked by the application.
They run tests of the web application to capture all the SQL queries in an allowlist and then enforce it at runtime.

Finally, Anna et al.~\cite{simpson_runtime_2013} propose an approach that uses dynamic analysis to create an allowlist of source code and associated file hashes.
At runtime, they get the source code of the assembly instruction being executed and verify it against the allowlist.
Then they use dynamic taint tracking to enforce the list at runtime.

\toolname can be considered as an integrity measurement based system, specialized for Java where the measurement happens while running the test suite of the application and then enforced at runtime.
Thus, it completely removes the manual work of verifying the integrity by client.
\toolname is novel in working at the granularity of dependencies, with an automated process of verifying the integrity of the application at runtime and stopping the process when the application loads a malicious class.
Such a high level of granularity helps developers to understand the reasons behind integrity violations.

\subsection{Other Kinds of Runtime Integrity}

Control flow integrity is about checking that the application executes according to the control flow graph as intended.
The survey by Burrow et al.~\cite{burow_controlflow_2017} provides a comprehensive overview of recent related work on control flow integrity for runtime protection.
\toolname is different from control flow integrity: 1) it is at a higher level of abstraction: dependencies; 2) it protects against a different class of attacks (see \autoref{sec:threat-model}).

Oblivious Hashing~\cite{ahmadvand_practical_2018} is a technique where the side effects of executed code are verified.
The hashing function instruments the code to add guards at the end of every control flow path.
At runtime, the guards ensure that the code is executing as expected by verifying the hash of the control flow path.
\toolname focuses on the prevention of execution of malicious classes rather than verifying the resulting state of the application.
Moreover, \toolname does not require any source code modification in the application whose runtime integrity needs to be preserved.

Deserialization attacks are common in Java applications~\cite{sayar_indepth_2023}.
In this class of attacks, the attacker sends a serialized object to the application and expects the application to deserialize it.
Upon deserialization, the malicious code is executed.
Filtering mechanisms~\cite{riggs_jep_2016} have been proposed which allow and deny certain classes to be deserialized.
However, Sayar et al.~\cite{sayar_indepth_2023} find that these mechanisms make the allowlist contain all possible serializable classes and the deny list is empty.
A solution proposed by the authors is to allowlist only the used classes of the application and \toolname is exactly on those lines.

\section{Conclusion}

In this paper, we proposed \toolname, a novel approach to prevent software supply chain attacks that exploit the dynamic class loading capabilities in Java.
\toolname builds a comprehensive Bill of Materials Index (\bomi) which includes all classes that are expected to be loaded at runtime.
In production, if the application loads a class that is not part of the \bomi, then \toolname terminates the application to prevent the injection of malicious code.
We demonstrate the effectiveness of \toolname by stopping 3 high-profile software supply chain attacks in Java libraries - \texttt{Log4j}, \texttt{H2}, and \texttt{Apache Commons Configuration}.
We also demonstrate that the construction and monitoring of a \bomi is compatible with real-world applications like \texttt{PDFBox}, \texttt{ttorrent}, and \texttt{GraphHopper}, with limited overhead.

As future work, we plan to extend \ac{SBOM} Runtime Watchdog to detect hidden classes~\cite{chung_jep_2019}.
As of today, no technology in the Java ecosystem can intercept the loading of hidden classes, but Dynamic Code Indexing can be extended to capture them.
Finally, our approach has goals similar to GraalVM~\cite{noauthor_graalvm_nodate}.
Both techniques aim to know all the classes before the application starts executing.
However, GraalVM relies on static analysis and \toolname relies on dynamic analysis and a thorough comparison of both approaches is on the research agenda.

\section{Acknowledgment}
We acknowledge the work of Elias Lundell, a research assistant at KTH Royal Institute of Technology, who helped in the replication of \texttt{Log4Shell}.
This work was supported by the CHAINS project funded by Swedish Foundation for Strategic Research (SSF), as well as by the Wallenberg Autonomous Systems and Software Program (WASP) funded by the Knut and Alice Wallenberg Foundation.
The icons in the figure are from \url{https://www.flaticon.com/}.


%

\bibliographystyle{IEEEtran}
\bibliography{main}





\end{document}